\documentclass[preprint2]{aastex}

\usepackage{graphicx}
\usepackage{bm}      
\usepackage{times}
\usepackage{color}

\newcommand{\vB}{\mathbf{B}}

\newcommand{\vv}{\mathbf{v}}

\newcommand{\vj}{\mathbf{j}}

\newcommand{\vk}{\mathbf{k}}
\newcommand{\vomega}{\mbox{\boldmath $\omega$}}

\shorttitle{Nonlinear Behavior of a Non-helical Dynamo}
\shortauthors{Mininni, Ponty, Montgomery, Pinton, Politano, \& Pouquet}

\begin{document}

\title{NONLINEAR BEHAVIOR OF A NON-HELICAL DYNAMO}

\author{Pablo D. Mininni\altaffilmark{1}, Yannick Ponty\altaffilmark{2}, 
        David C. Montgomery\altaffilmark{3}, Jean-Francois Pinton
        \altaffilmark{4}, \\ 
        Helene Politano\altaffilmark{2}, and Annick Pouquet\altaffilmark{1}}
\altaffiltext{1}{National Center for Atmospheric Research, P.O. Box 3000, 
                 Boulder, Colorado 80307-3000}
\altaffiltext{2}{Laboratoire Cassiop\'ee, Observatoire de la C\^ote 
                 d'Azur, BP 4229, Nice Cedex 04, France}
\altaffiltext{3}{Dept. of Physics and Astronomy, Dartmouth College, 
                 Hanover, NH 03755}
\altaffiltext{4}{Laboratoire de Physique, CNRS \& \'Ecole Normale 
                 Sup\'erieure de Lyon, 46 All\'ee d'Italie, 69007 Lyon, 
                 France}

\begin{abstract}
A three-dimensional numerical computation of magnetohydrodynamic 
dynamo behavior is described. The dynamo is mechanically forced 
with a driving term of the Taylor-Green type. The magnetic field 
development is followed from negligibly small levels to saturated 
values that occur at magnetic energies comparable to the kinetic 
energies. Though there is locally a helicity density, there is 
no overall integrated helicity in the system. Persistent oscillations 
are observed in the saturated state for not-too-large mechanical 
Reynolds numbers, oscillations in which the kinetic and magnetic 
energies vary out of phase but with no reversal of the magnetic 
field. The flow pattern exhibits considerable geometrical structure 
in this regime. As the Reynolds number is raised, the oscillations 
disappear and the energies become more nearly stationary, but retain 
some unsystematically fluctuating turbulent time dependence. The 
regular geometrical structure of the fields gives way to a more 
spatially disordered distribution. The injection and dissipation 
scales are identified and the different components of energy transfer 
in Fourier space are analyzed, in particular in the context of 
clarifying the role played by different flow scales in the 
amplification of the magnetic field.
\end{abstract}


\section{\label{sec:intro}INTRODUCTION}

Evidence of the existence of magnetic fields is known in many 
astronomical objects. These fields are believed to be generated and 
sustained by a dynamo process (e.g. \citet{Moffatt}), and often these 
objects are characterized by the presence of large scale flows (such 
as rotation) and turbulent fluctuations. These two ingredients are known 
to be crucial for magnetohydrodynamic dynamos. In recent years, significant 
advances have been made either studying large scale flows dynamos in the 
kinematic approximation, or using direct numerical simulations to study 
turbulent amplification of magnetic fields in simplified geometries.

In a previous paper \citep{Ponty04}, a study of the self-generation 
of magnetic fields in a turbulent conducting fluid was reported. The 
study was computational and dealt mainly with effects of lowering 
the magnetic Prandtl number $P_M$ of the fluid (ratio of kinematic 
viscosity to magnetic diffusivity). The velocity field was externally 
excited by a forcing term on the right hand side of the equation of 
motion whose geometry was that of what has come to be called the 
``Taylor-Green vortex'' 
\citep{Taylor37,Morf80,Pelz85,Nore97,Marie03,Bourgoin04}. The regime of 
operation was one of kinetic Reynolds number $\gg 1$ (so that the 
fluid motions were turbulent), and the emphasis was on how large the 
magnetic Reynolds numbers had to be in order that infinitesimal magnetic 
fields could be amplified and grow to macroscopic values.

Here, we want to describe and stress another aspect of the 
Taylor-Green dynamo. In particular, we have found computationally 
that it has an oscillatory regime, for not too large a Reynolds number, 
in which energy is passed back and forth regularly between the 
mechanical motions and the magnetic excitations in a way we believe 
to be new. Out of the velocity field emerges a geometrically-regular, 
time-averaged pattern involving coherent magnetic and mechanical 
oscillations.

As the Reynolds number is increased, the resulting flow has 
a well defined large scale pattern and non-helical turbulent 
fluctuations. In this case, the oscillations disappear and the 
magnetic field grows at scales both larger and smaller than the 
integral scale of the flow. After the nonlinear saturation of the 
dynamo, velocity field fluctuations are partially suppressed and a 
magnetic field with a spatial pattern reminiscent of the low Reynolds 
number case can be identified. This complex evolution of the magnetic 
field can be understood studying the role played by the energy transfer 
in Fourier space.

In Sec. \ref{sec:computation}, we describe the numerical experiments 
and outline a typical time history of the development of an 
oscillatory dynamo. We then go on to show how, by increasing the 
Reynolds number, the oscillatory behavior can be suppressed. In 
Sec. \ref{sec:results}, we make use of color displays of the field 
quantities to demonstrate the cycle of the oscillation and to reveal 
the intriguing and complexly varying three-dimensional pattern that 
characterizes it. The pattern, though regular, is difficult to see 
through completely in physical terms. Finally, Sec. \ref{sec:discussion} 
suggests some precedents, provides a partial explanation, and 
considers other similar situations where such coherence may or may 
not be expected to emerge out of turbulent disorder.

\section{\label{sec:computation}THE COMPUTATION}

The Taylor-Green vortex is a flow with an initial periodic velocity 
field
\begin{equation} 
\vv_{\rm TG}(k_0)= \left[ \begin{array}{c} 
     \sin(k_0 x) \cos(k_0 y) \cos(k_0 z)  \\ 
     -\cos(k_0 x) \sin(k_0 y) \cos(k_0 z) \\ 
     0  \end{array} \right] ,
\label{eq:TG}
\end{equation} 
and was originally motivated as an initial condition that, though 
highly symmetric, would lead to the rapid development of small 
spatial scales \citep{Taylor37}. We introduce it here on the right 
hand side of the magnetohydrodynamic (MHD) equation on motion for 
the velocity field $\vv$:
\begin{equation}
\frac{\partial \vv}{\partial t} + \vv \cdot \nabla \vv = -\nabla {\cal P} 
     + \vj \times \vB - \nu \nabla \times \vomega + F \vv_{\rm TG} , 
\label{eq:NS}
\end{equation}
where $\vB$ is the magnetic field, advanced by
\begin{equation}
\frac{\partial \vB}{\partial t} + \vv \cdot \nabla \vB = 
     \vB \cdot \nabla \vv - \eta \nabla \times \vj . 
\label{eq:ind}
\end{equation}
Eqs. (\ref{eq:NS}) and (\ref{eq:ind}) are to be solved pseudospectrally. 
The current density is $\vj = \nabla \times \vB$ (we use the common 
dimensionless ``Alfv\'enic'' units), $F$ is a forcing amplitude, and 
$k_0 = 2 (2 \pi/L)$, where $L=2 \pi$ is chosen as the basic periodicity 
length in all three directions. In the incompressible case, 
$\nabla \cdot \vv = 0$ and $\nabla \cdot \vB = 0$; $\nu^{-1}$ and 
$\eta^{-1}$ are (dimensionless) mechanical and magnetic Reynolds numbers 
since we take as characteristic velocity and length $U_0 = 1$, $L_0 = 1$ 
leading to an eddy turnover time of order unity; and ${\cal P}$ is the 
dimensionless pressure, normalized by the (uniform) mass density.

The strategy is to turn on a non-zero force $F$ at $t=0$ and allow the 
code to run for a time as a purely Navier-Stokes code, with the $\vB$ 
and $\vj$ fields set at zero. The initial velocity field is given by 
\begin{equation}
\vv_0 = -\frac{F}{\nu} \nabla^{-2} \vv_{\rm TG} .
\end{equation}
and the amplitude of $F$ is set to obtain an initial unitary 
{\it r.m.s.} velocity. As the system evolves, more modes are excited 
and the dissipation increases. To maintain the kinetic energy at the 
same level, the amplitude of the force is controlled during the 
hydrodynamic simulation to compensate the dissipation. At each time $t$, 
the energy injection rate
\begin{equation}
\epsilon = F(t) \int{ \vv \cdot \vv_{\rm TG} \, \textrm{d}^3x} ,
\end{equation}
and the enstrophy
\begin{equation}
\Omega = \int{ \left( \nabla \times \vv \right)^2 \, \textrm{d}^3x} ,
\end{equation}
are computed, and the amplitude of the external force needed to overcome 
dissipation is computed as
\begin{equation}
F^* = \frac{2 \nu \Omega F(t)}{\epsilon}
\end{equation}
The response of the velocity field to the change in the external force 
has a certain delay, and to avoid spurious fluctuations the average 
value $\left< F^* \right>$ of this quantity is computed for the last 
nine time steps, as well as the averaged error in the energy balance 
${\cal E} = \left< 2 \nu \Omega - \epsilon \right>$. Finally, the 
amplitude of the external force at time $t+\Delta t$ is updated as
\begin{equation}
F(t+\Delta t) = 0.9 F^* + \left( 0.1 \left< F^* \right> + 0.01 \cal{E} 
    \right)/9 .
\end{equation}
Once a stationary state is reached, the last computed amplitude of 
the force can be used to restart the simulation with constant force 
instead of constant energy. In this case, the energy fluctuates 
around its original value, and the {\it r.m.s} velocity averaged 
in time is unitary. This value of the {\it r.m.s} velocity, and the 
integral length scale $\ell_{int}$ of the resulting flow are used 
to defined the Reynolds numbers in the following sections. For a 
different scheme to compensate the dissipation see e.g. 
\citet{Archontis03}. 

Once the stationary kinetic state is reached, the magnetic field is 
seeded with randomly-chosen Fourier coefficients and allowed to 
amplify. All the magnetohydrodynamic simulations are done with 
constant force, and the amplitude $F$ is obtained as previously 
discussed. The initial magnetic field is non-helical, with the 
magnetic energy smaller than the kinetic energy at all wavenumbers, 
and a spectrum satisfying a $k^2$ power law at large scales and an 
exponential decay at small scales. A previous paper has described 
the ``kinematic dynamo'' regime in which the magnetic excitations, 
while growing, are too small to affect the velocity field yet 
\citep{Ponty04}. In particular, a threshold curve for magnetic field 
amplification was constructed in the plane whose axes are magnetic 
Prandtl number, $P_M \equiv \nu / \eta$, and magnetic Reynolds number. 
As $P_M^{-1}$ increases, there is a sharp rise in the dynamo threshold, 
followed by a plateau. Here, the purpose is to follow the evolution 
of $\vB$ out of the kinematic regime and observe whatever saturation 
mechanisms may set in.

\section{\label{sec:results}COMPUTATIONAL RESULTS}

\begin{figure}
\plotone{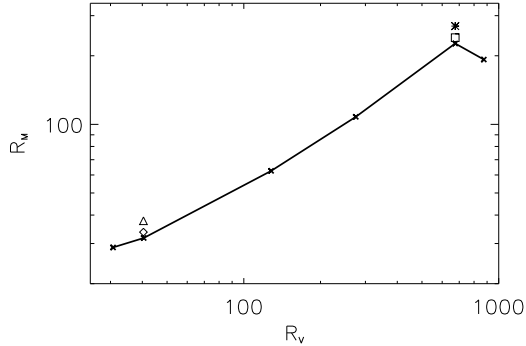}
\caption{Critical magnetic Reynolds $R_M^c$ for dynamo action (thick solid 
         line) inferred from direct numerical simulations (crosses), as a 
         function of $R_V$ \citep{Ponty04}. The position of the runs 
         discussed in Table \ref{table} are indicated by symbols: run A 
         (diamond), A$'$ (triangle), B (square), and B$'$ (star).}
\label{fig:RMC}
\end{figure}

Table 1 summarizes the parameters of the four runs we have carried 
out. Runs A and A$'$ have relatively low mechanical and magnetic 
Reynolds numbers ($\sim 40$, based on the integral length scale 
and the {\it r.m.s} velocity) while runs B and B$'$ have mechanical 
Reynolds numbers $R_V = 675$. The magnetic Reynolds numbers $R_M$ 
for runs A, A$'$ are $33.7$ and $37.8$, respectively, while those for 
B, B$'$ were $240.2$ and $270$, respectively. These values of $R_M$ 
were in all cases above the previously-determined thresholds 
\citep{Ponty04} for magnetic field growth (see Fig. \ref{fig:RMC}). 
Note that $R_M$ for runs A and B is 6\% above threshold, while 
runs A$'$ and B$'$ are 20\% above threshold. We chose $k_0=2$ in all 
cases, so that the kinetic energy spectrum peaks at 
$k = k_0 \sqrt{3} \approx 3$. As previously mentioned, the amplitude 
of the external force was constant during the MHD simulation, and 
given by $F = 0.926$ in runs A and A$'$, and $F = 0.37$ in runs 
B and B$'$.

\begin{table*}[!h]
\begin{center}
\begin{tabular}{|c|c|c|c|c|c|c|c|c|c|c|}
\hline \hline
Run & $\nu$         & $\eta$        & $R_V$ & $R_M$ & $R_M^c$ & $P_M$ 
& $\ell_{int}$ & $\lambda$ & $N$     & $\Delta$ \\
\hline \hline
A   & $5\; 10^{-2}$ & $6\; 10^{-2}$ & 40.5  & 33.7  & 31.7    & 0.83  
& 2.02       & 1.69      & $64^3$  & 6        \\
\hline
B  & $2\; 10^{-3}$ &$5.62\; 10^{-3}$& 675   & 240.2 & 226.4   & 0.35  
& 1.35       & 0.6       & $256^3$ & 6        \\
\hline
A$'$& $5\; 10^{-2}$ &$5.35\; 10^{-2}$& 40.5 & 37.8  & 31.7    & 0.93  
& 2.02       & 1.69      & $64^3$  & 20       \\
\hline
B$'$& $2\; 10^{-3}$ & $5\; 10^{-3}$ & 675   & 270   & 226.4   & 0.4   
& 1.35       & 0.6       & $256^3$ & 20       \\
\hline \hline
\end{tabular}
\end{center}
\label{table}
\caption{Simulations. The Reynolds numbers $R_V$ and $R_M$ are based on 
         the integral lengthscale $\ell_{int}$, while $\lambda$ is the 
         Taylor scale. $R_M^c$ is the critical Reynolds number obtained 
         for $R_V$, and the magnetic Prandtl number is $P_M = \nu/\eta$. 
         $N$ is the spatial grid resolution used in the simulation, and 
         $\Delta$ gives the percentage above threshold in $R_M$ for 
         each simulation.}
\end{table*}

\subsection{Low Reynolds numbers and close to threshold}

\begin{figure}
\plotone{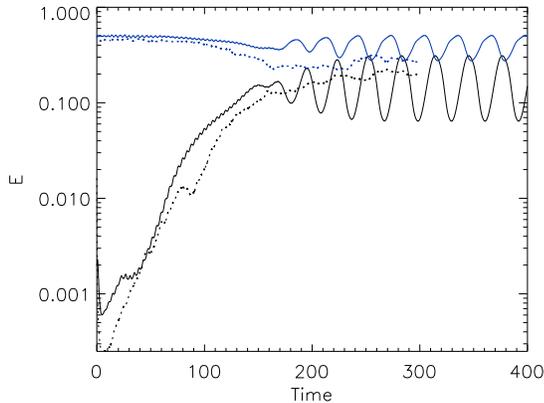}
\caption{(color online) Time history of kinetic (upper blue curves) and 
         magnetic energy (lower curves) for runs A at low Reynolds 
         (solid lines) and B at high Reynolds (dotted lines) both 6\% 
         above threshold.}
\label{fig:EAB}
\end{figure}

The behavior at saturation is very different for the high and low 
Reynolds numbers. The histories of the energies for runs A and B (both 
6\% above threshold) are displayed in Fig. \ref{fig:EAB}. The upper two 
curves are the kinetic energies of these runs, a solid line for run 
A and a dotted one for run B. The lower two curves are the magnetic 
energies, with the same conventions. The origin of time is chosen 
from the moment when the seed magnetic fields are introduced.

It is clear that saturation is achieved unsystematically for the 
high $R_V$ run B, with the resulting magnetic energy smaller than 
the kinetic energy and both in a statistically steady state. The 
solid lines associated with the lower Reynolds number 
run A, however, show a systematic, sharp oscillation in both energies, 
with the maxima of one coinciding with the minima of the other. This 
is clearly a significantly different behavior from the high $R_V$ 
case, and is only partially understood. Such out-of-phase oscillations 
have already been observed in the nonlinear regime in constrained 
geometries, for example using a quasi-geostrophic model for strongly 
rotating flows \citep{Schaeffer04}, or a 2.5D formulation for the Ekman 
layer instability \citep{Ponty01}.

In the two simulations presented in Fig. \ref{fig:EAB} $R_M$ is 6\% 
larger than $R_M^c$, and the growth rates during the kinematic regime 
are similar. While $\eta$ is one order of magnitude smaller in run B 
than in run A, the nonlinear saturation in both runs takes place at 
approximately the same time. In both runs the integral turnover time 
is approximately the same. This contrasts with dynamos in flows with 
net helicity, where the nonlinear saturation was shown to occur in a 
magnetic diffusion time \citep{Brandenburg01a} (this diffusion time 
is of the order of $50$ for run A, and $560$ for run B). Note that 
although the flow generated by the Taylor-Green force is locally 
helical, the net helicity of the flow in the entire domain is zero.

The forcing term generating the flow from Eqs. (\ref{eq:NS}) and 
(\ref{eq:ind}) is initially entirely in the horizontal ($x,y$) 
directions. It is essentially a vortical flow whose phase oscillates 
with increasing $z$. The velocity field in Eq. (\ref{eq:NS}) is 
not, however, a steady state, and vertical ($z$) components 
develop quickly, leading to an approximately meridional flow to 
be added to the toroidal one in each cell. A total streamline will 
resemble the shape of a wire wrapped around the outside of a 
doughnut, diagonally, which enters the hole of the doughnut 
at the bottom and emerges at the top.

The amplification process for the magnetic field is difficult to 
visualize in this geometry. Field lines seem to be sucked into the 
hole of the doughnut and stretched and twisted in the process. The 
resulting amplified magnetic flux is then deposited and piled up 
in the horizontal planes between the cells. This flux, in turn, is 
the source of the field lines which are further sucked into the 
holes in the doughnut and amplified. In the kinematic regime, 
but in a different geometry (including boundary conditions), the 
amplification of a magnetic field by a similar flow was also 
discussed in \citet{Marie03,Bourgoin04}.

Throughout the process, the rate of doing work by the magnetic 
field on the velocity field originates in the Lorentz force 
contribution, $-(\vj \times \vB) \cdot \vv$. This energy input 
into the magnetic field is Ohmically dissipated by the $\eta j^2$ 
integral. As the magnetic field grows, the fluid must work harder 
mechanically, because $\vj$ and $\vB$ are increasing. Since $F$ is 
constant, eventually a limit is reached where $\vv$ can no longer 
transfer energy to $\vB$ at its previous rate and slows down. 
At that point, the magnetic energy begins to be transferred in 
the reverse sense so that $\vv$ grows again as $\vj$ and $\vB$ 
become weaker. The cyclic nature of the process ensues.

\begin{figure}
\plotone{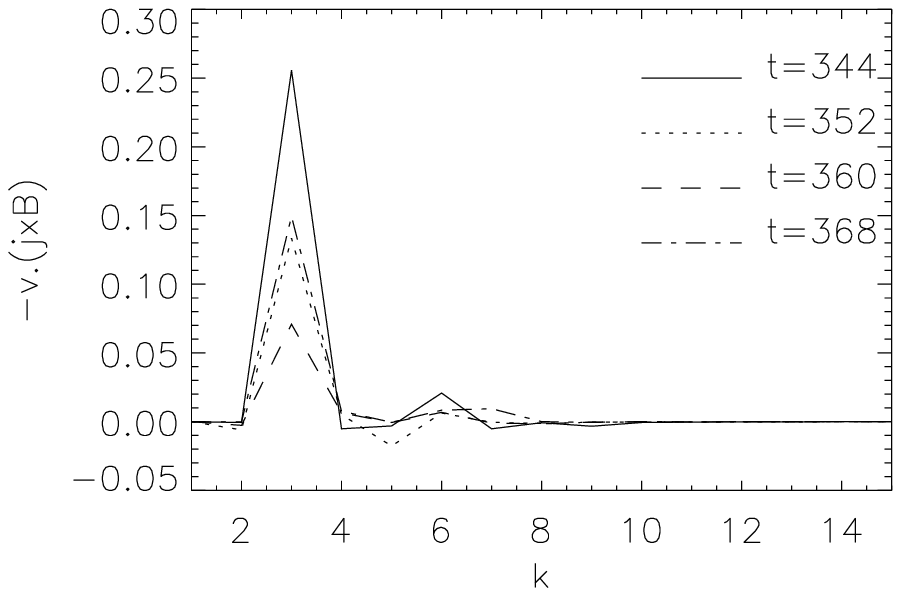}
\caption{Fourier spectrum of $-(\vj \times \vB) \cdot \vv$ for 
         different times for run A for one oscillation.}
\label{fig:jtrans_lowRe}
\end{figure}

It is revealing to decompose $(\vj \times \vB) \cdot \vv$ spectrally 
and plot the Fourier transform 
$-\widehat{[(\vj \times \vB) \cdot \vv]}_\vk$ as a function of $k$, 
as shown in Fig. \ref{fig:jtrans_lowRe} for run A. The peak near $k=3$ 
shows that this is the region where the mechanical work is being done to 
create the magnetic energy. The curve is plotted at four times during 
a complete oscillation, including $t=344$, where the magnetic energy 
is at its maximum during the cycle, and $t=360$, where the magnetic 
energy is at its minimum.

\begin{figure}
\plotone{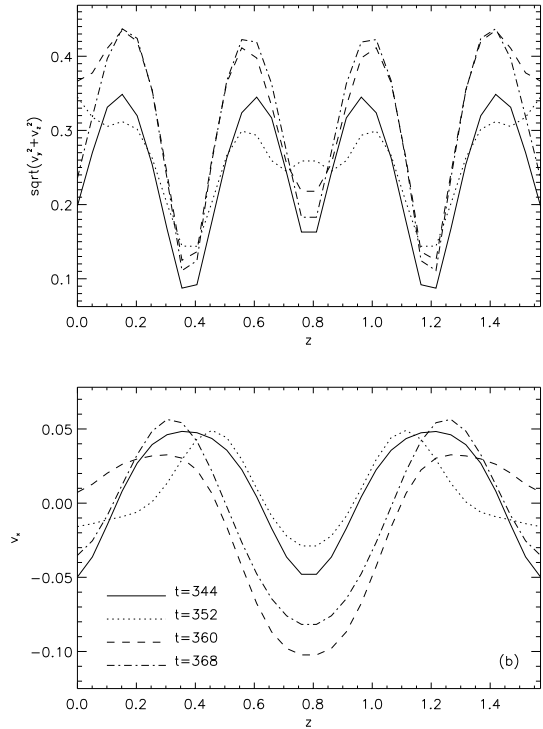}
\caption{(a) $(v_y^2+v_z^2)^{1/2}$, and (b) $v_x$ as a function of $z$ 
         at $x=3\pi/8$ and $y=\pi/4$, and at different times for 
         run A, for one oscillation.}
\label{fig:meridional}
\end{figure}

\begin{figure*}
\includegraphics[width=15cm]{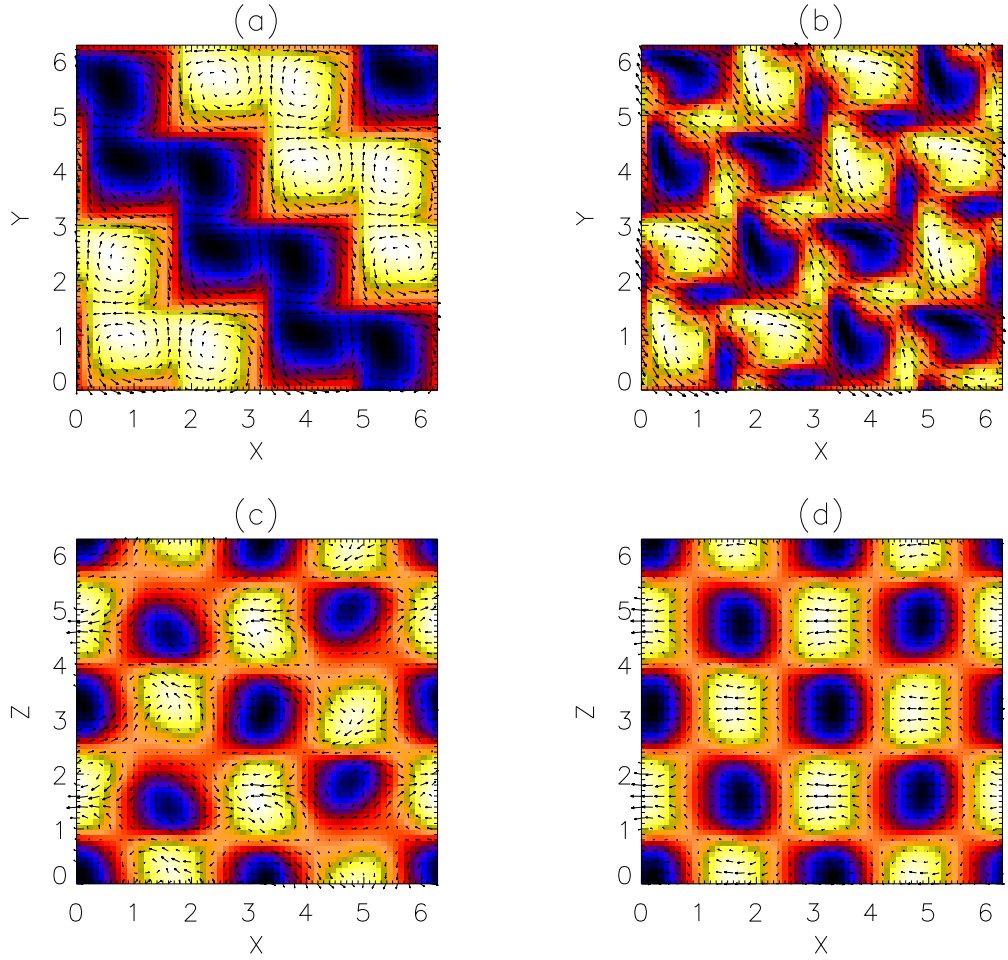}
\caption{(color online) Upper panels: cross-sectional plots of the 
         velocity field in the plane $z=0$ at (a) $t=352$, and (b) 
         $t=368$, for run A. The arrows show the directions of $v_x$, 
         $v_y$ and the colors indicate the values of $v_z$, positive 
         (light) or negative (dark). Lower panels: cross-sectional plots 
         of the velocity field in the plane $y=\pi/4$ at (c) $t=352$, 
         and (d) $t=368$ ($v_x$, $v_z$ indicated by the arrows, and 
         $v_y$ by the color).}
\label{fig:vcut}
\end{figure*}

There is considerable structure to the flow for these low Reynolds 
number cases, anchored by the driving term in Eq. (\ref{eq:NS}). 
Figs. \ref{fig:meridional} show instantaneous plots of the velocity 
field components along a vertical cut at $x=3\pi/8$ and $y=\pi/4$, 
as functions of $z$ for run A. This cut corresponds to a line in the 
$z$ direction displaced (in the $xy$ plane) out of the center line of 
the vortices imposed by the external Taylor-Green force (corresponding 
to $x=y=\pi/4$). Plotted in Fig. \ref{fig:meridional}.a is 
$(v_y^2+v_z^2)^{1/2}$ {\it vs.} $z$, and in Fig. \ref{fig:meridional}.b, 
$v_x$ {\it vs.} $z$. In both curves, four different times are shown. 
In this cut, $v_x$ corresponds to the amplitude of the toroidal 
flow associated with the vortices imposed by the forcing, while 
$(v_y^2+v_z^2)^{1/2}$ can be associated with the meridional flow 
previously defined. Note the mirror symmetries satisfied by the flow. 
As the oscillations evolve, not only the amplitude of the flow changes, 
but also the position of the maxima are slightly displaced. The flow 
geometry will be clarified in more detail in Figs. \ref{fig:vcut}.

In Figs. \ref{fig:vcut}.a,b are exhibited cross-sectional plots of the 
velocity field in the plane $z=0$ at two different times for run A. The 
arrows show the directions of $v_x$, $v_y$ and the colors indicate the 
values of $v_z$, positive (light) or negative (dark), at the same 
locations. Note the sixteen vortices imposed by the external Taylor-Green 
forcing with $k_0 = 2$. The amplitude of these vortices is modulated 
in $z$, and at $\pi/2$ the same structure is obtained in the flow but 
with the vortices rotating in the opposite direction. Most of the 
stretching of magnetic field takes place in these cells. Between these 
structures, at $z=\pi/4$, stagnation points are present where the 
magnetic field piles up, as will be shown.

Figs. \ref{fig:vcut}.c,d display similar plots at the plane $y=\pi/4$, 
with this time $v_x$, $v_z$ indicated by the arrows, and $v_y$ by the 
color. The regions of alternating color correspond to the 
cross-section of the vortices imposed by the Taylor-Green forcing. 
Also the meridional flow can be identified in these cross-sections. 
However, note that this flow during the cycle is modified by the 
magnetic field in a more dramatic way than the toroidal flow. As 
previously shown in Figs. \ref{fig:jtrans_lowRe} and 
\ref{fig:meridional}, the Lorentz force mostly opposes the velocity 
field at large scales. The final effect of the magnetic field on the 
flow seems to be to suppress small-scale fluctuations, leaving a 
well-ordered pattern. This effect will be more dramatic at large 
$R_V$, as will be shown in the next subsection.

\begin{figure*}
\includegraphics[width=15cm]{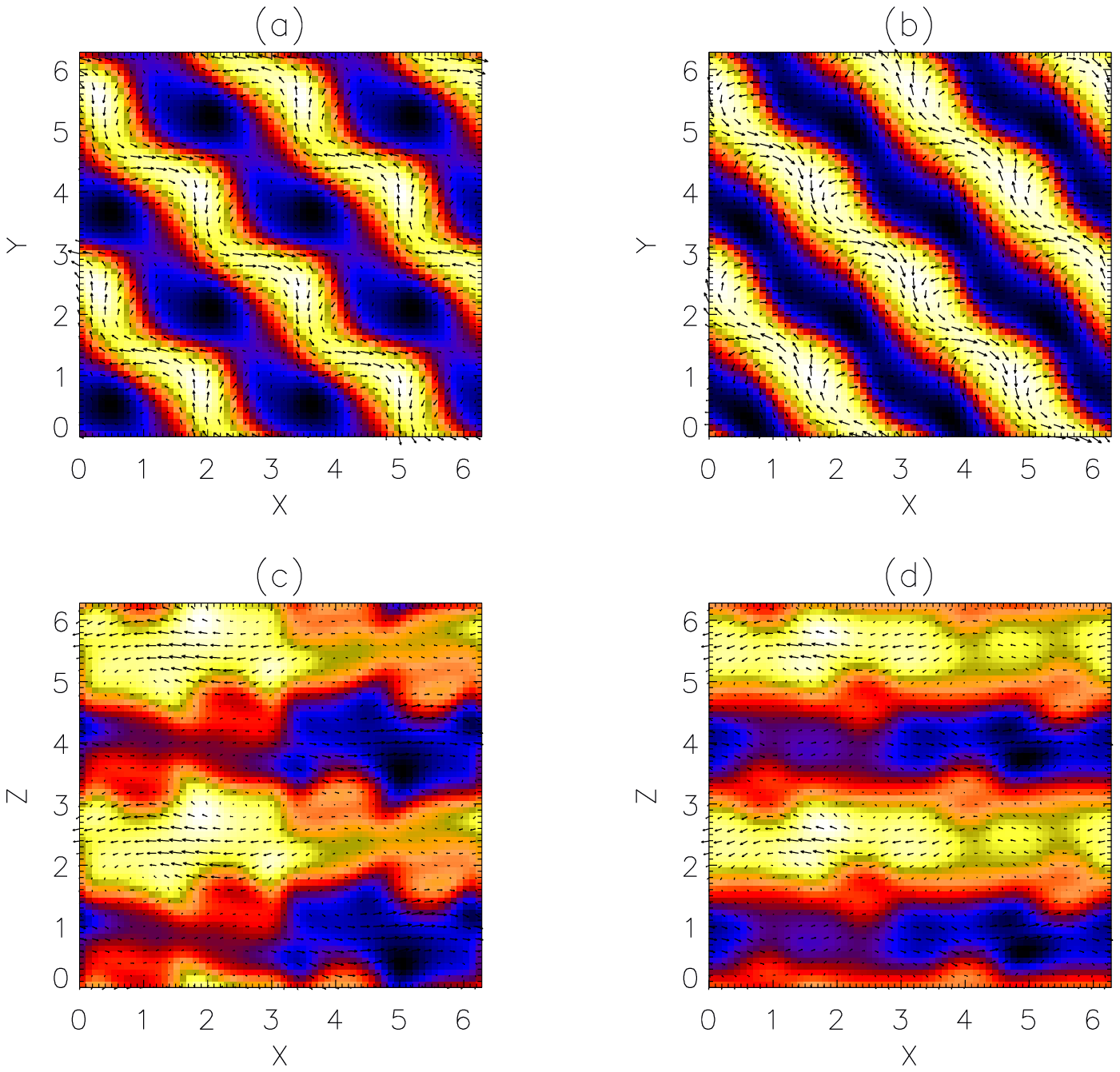}
\caption{(color online) Upper panels: cross-sectional plots of the 
         magnetic field in the plane $z=0$ at (a) $t=352$, and (b) 
         $t=368$, for run A. Lower panels: cross-sectional plots of the 
         magnetic field in the plane $y=\pi/4$ at (c) $t=352$, and (d) 
         $t=368$. Same conventions than in Fig. \ref{fig:vcut}.}
\label{fig:bcut}
\end{figure*}

Figs. \ref{fig:bcut}.a,b show the magnetic field in the plane 
$z = 0$ at the same times with the same conventions ($B_x$, $B_y$ 
are arrows, $B_z$ is indicated by color), again for run A. The 
stretching of magnetic field lines by the toroidal flow can be 
observed in these sections. Figs. \ref{fig:bcut}c.d show the 
magnetic field in the plane $y=\pi/4$ at the same times, and with 
the same plotting conventions. Note in dark and light colors the 
horizontal bars where most of the magnetic energy is concentrated. 
These regions correspond to stagnation planes of the external 
Taylor-Green forcing.

\begin{figure*}
\includegraphics[width=15cm]{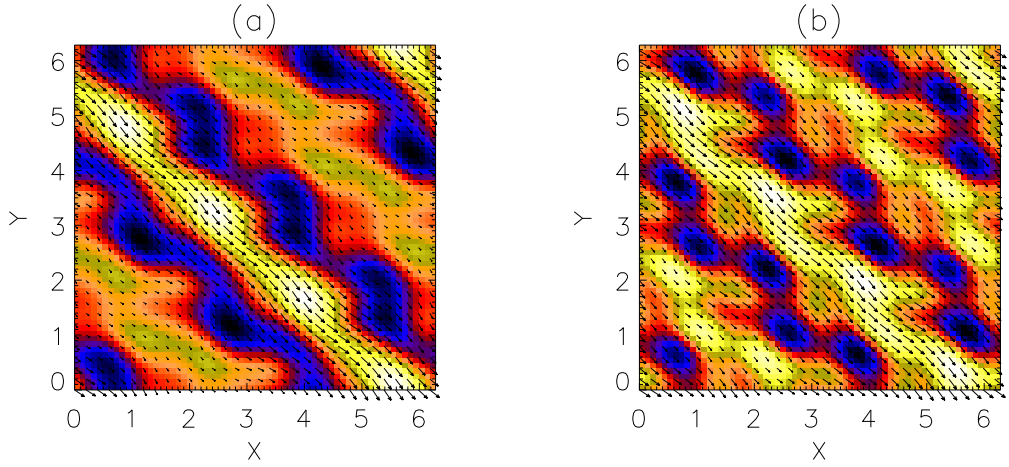}
\caption{(color online) cross-sectional plots of the 
         magnetic field in the plane $z=\pi/4$ at (a) $t=352$, and (b) 
         $t=368$, for run A. Same conventions than in Fig. 
         \ref{fig:vcut}.}
\label{fig:bsta}
\end{figure*}

Fig. \ref{fig:bsta}, finally, shows the magnetic field in the plane $z=0$ 
at different times for run A. This is a plane between rows of basic cells 
and is a candidate where the amplified flux ``piles up'' as previously 
indicated. It is apparent that the dynamical variation is much less in 
this plane during the cycle. Also, in Fig. \ref{fig:bsta} note the presence 
of locally ``dipolar'' structures (light and dark regions), centered in 
each of the Taylor-Green cells. These structures correspond to the almost 
uniform (and mostly concentrated in the $x$,$y$ plane) magnetic field 
being sucked into the hole of the doughnut given by the Taylor-Green 
force.

A more detailed picture of the dynamics of the forced Taylor-Green 
dynamo at low Reynolds numbers has eluded us, but it is imaginable that 
in less complex flows a more comprehensive understanding of the low 
$R_V$ non-helical dynamo may be possible.

\subsection{High Reynolds numbers and further from threshold}

\begin{figure}
\plotone{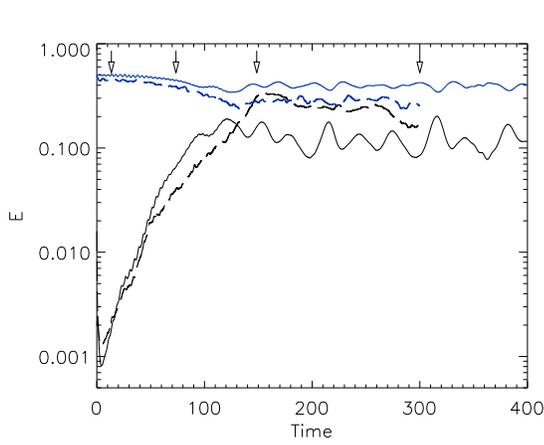}
\caption{(color online) Time history of kinetic (upper blue curves) and 
         magnetic energy (lower curves) for runs A$'$ at low Reynolds 
         (solid lines) and B$'$ at high Reynolds (dotted lines), both 
         20\% above threshold. The arrows on the top of the figure 
         represent the times at which the transfer terms displayed in 
         Fig. \ref{fig:transfer} are evaluated.}
\label{fig:Eprimes}
\end{figure}

Runs B, B$'$ involve higher Reynolds numbers and behave rather differently 
from runs A, A$'$. Fig. \ref{fig:Eprimes} contrasts the time histories 
of the kinetic energies and magnetic energies for run A$'$ (solid curves) 
and run B$'$ (dashed), both 20\% above threshold. The upper curves are 
kinetic energies and the lower curves are magnetic energies. It is clear 
that the B$'$ runs saturate at a level of near-equipartition and do not 
exhibit the oscillatory behavior seen in the lower Reynolds number runs. 
Run A$'$ retains a vestige of the periodic behavior, seen most clearly 
in the magnetic energy curve which is quasi-periodic or close to ``chaotic''. 
Note the overshooting of the magnetic energy for run B$'$ near 
$t \sim 150$, linked to the large drop in kinetic energy. Note also the 
similar growth rates (as for runs A and B) although the magnetic 
diffusivities differ again by almost an order of magnitude.

\begin{figure}
\plotone{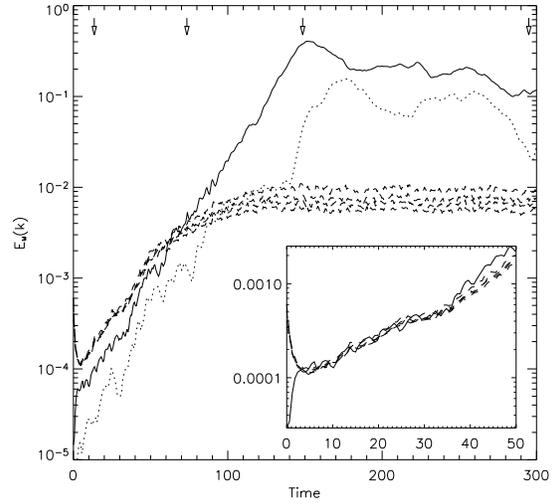}
\caption{$B^2$ integrated over different shells in Fourier space as a 
         function of time, for run B$'$ and $k=1$ (dotted line), $k=2$ 
         (solid) and $k=9,10,11,12$ (dashed lines). The inset shows 
         the evolution at early times, with all the shells normalized 
         to have the same amplitude. The arrows are at the same times 
         as in Fig. \ref{fig:Eprimes}.}
\label{fig:Emodes}
\end{figure}

During the exponential period of the magnetic energy growth, it is 
of interest to note that the various Fourier modes all appear to be 
growing at the same rate in run B$'$ (the same effect is observed 
in run B). This can be seen by separating the Fourier space into 
``shells'' of modes of the same width $\Delta k$. The time histories 
of these shells are plotted in Fig. \ref{fig:Emodes}. In the inset, 
all the shells have been normalized to have the same amplitudes 
per $k$-mode at $t=4$, to show that the exponentiation rates up to 
about $t=30$ are the same or nearly so. This behavior is 
characteristic of small scale dynamos \citep{Kazantsev67,Brandenburg01a}. 
Note that after $t=30$ the shell with $k=2$ seems to start growing 
faster than the small scale modes. Shortly after this time, the small 
scales saturate and the large scale magnetic field keeps growing 
exponentially up to $t=150$.

\begin{figure}
\plotone{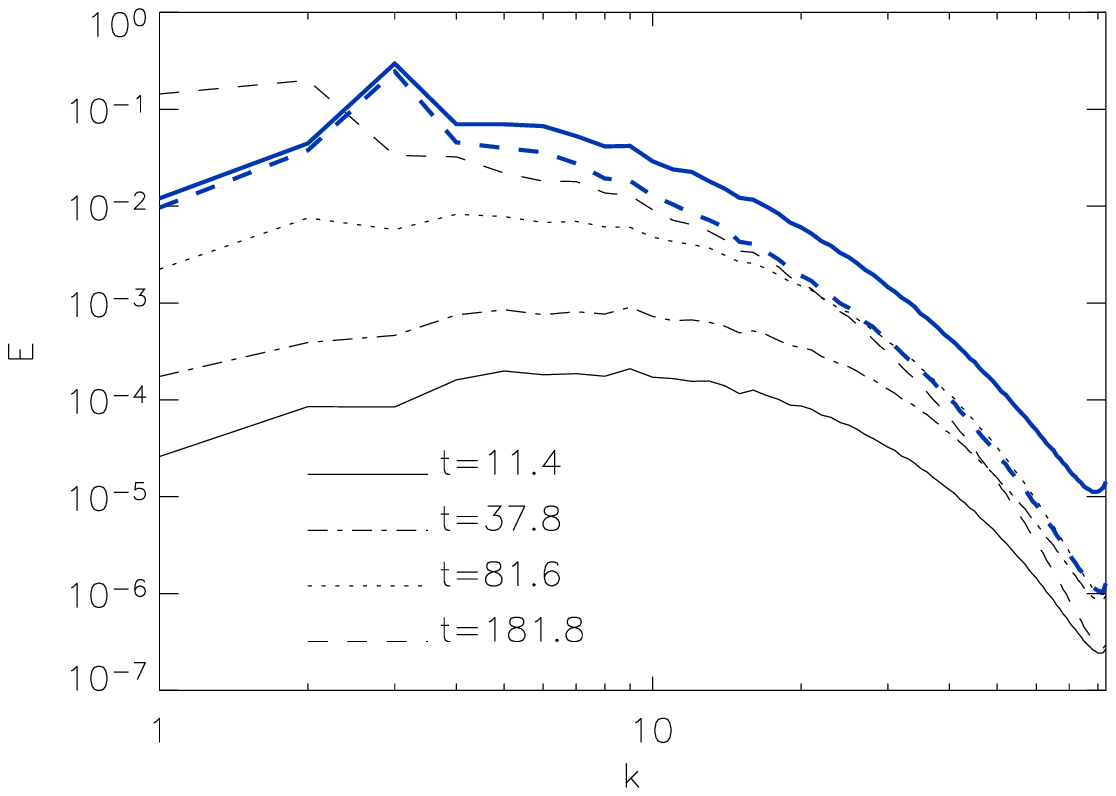}
\caption{(color online) Kinetic (thick blue lines) and magnetic 
         energy spectra (thin lines) as a function of time for run B$'$. 
         Kinetic spectra are only shown at $t=11.4$ and $t=181.8$; 
         note in the latter case, the strong diminution at small 
         scales of the velocity toward magnetic scales.}
\label{fig:Bspectra}
\end{figure}

The total kinetic energy spectra (thick lines) and magnetic energy 
spectra (thin lines) for run B$'$ are shown in Fig. \ref{fig:Bspectra}. 
Only two kinetic spectra are shown, at times $t=11.4$ and $t=181.8$. 
At early times, the magnetic energy spectrum peaks at small 
scales ($k \approx 9$), and the spectrum at large scales seems to 
satisfy a $k^{3/2}$ power law as already observed for the Taylor-Green 
flow \citep{Ponty04}, and for other flows as well \citep{Haugen04}. 
The magnetic energy increases from $t=11.4$ to $t=181.8$ and eventually 
dominates the kinetic energy at the longest wavelength.

The appearance of these quasi-dc components of the magnetic field seems 
to have a profound effect on the short-wavelength kinetic spectral 
components, depressing them by an order of magnitude, as is also 
visible from the thick broken line in Fig. \ref{fig:Bspectra}. The 
most straightforward interpretation is in terms of what is sometimes 
called the ``Alfv\'en effect''. The idea is that in incompressible 
MHD, any nearly spatially uniform, slowly varying, magnetic field 
forces the small scale excitations to behave like Alfv\'en waves. 
In an Alfv\'en wave, the energy is generally equipartitioned 
between magnetic field and velocity field, and any mechanism which 
damps one will damp the other. Since $\eta \gg \nu$ when $P_M \ll 1$, 
the Kolmogorov ``inner scale'' can be defined entirely in terms 
of energy dissipation rate and $\eta$, regardless of how much smaller 
the viscosity is. This was already observed in closure computations 
of MHD turbulence of low $P_M$ \cite{Leorat81}.

One could jump to the conclusion that for $\nu/\eta \ll 1$, the 
dynamo process will behave as if $P_M$ were of $\mathcal{O} (1)$ 
(see \citet{Yousef03} for different simulations supporting 
this conclusion). This is certainly inappropriate in the formation, 
or ``kinematic,'' phase, when the magnetic field is small but amplifying 
and there is no quasi-dc magnetic field to enforce the necessary 
approximate equipartition at small scales. This conclusion can 
also apply in more complex systems, such as during the reversals of the 
Earth's dynamo.

The central role played by the $-\vv \cdot (\vj \times \vB)$ term by 
which energy is injected into the magnetic field can be clarified by 
plotting the transfer functions $T(k)$ for the magnetic field and 
velocity field as functions of $k$ at different times.

The energy transfer function
\begin{equation}
T(k) = T_V(k)+T_M(k),
\label{tottran}
\end{equation}
represents the transfer of energy in $k$-space, and is obtained by 
dotting the Fourier transform of the nonlinear terms in the momentum 
equation (\ref{eq:NS}) and in the induction equation (\ref{eq:ind}), 
by the Fourier transform of $\vv$ and $\vB$ respectively. It also 
satisfies
\begin{equation}
0=\int_0^\infty{T(k')\,\textrm{d}k'},
\end{equation}
because of energy conservation by the non-linear terms; one can also 
define 
\begin{equation}
\Pi(k) = \int_0^k{T(k')\,\textrm{d}k'},
\end{equation}
where $\Pi(k)$ is the energy flux in Fourier space. In equation 
(\ref{tottran}), $T_V(k)$ is the transfer of kinetic energy
\begin{equation}
T_V(k) = \int{ \hat{\vv}_{\vk} \cdot \left[-\widehat{\left(\vomega \times 
    \vv\right)}_{\vk} +\widehat{\left(\vj \times 
    \vB\right)}_{\vk} \right]^* \,\textrm{d}\Omega_{\vk}},
\end{equation}
where the hat denotes Fourier transform, the asterisk complex conjugate, 
and $d\Omega_{\vk}$ denotes integration over angle in Fourier space. In this 
equation and the following, it is assumed that the complex conjugate part 
of the integral is added to obtain a real transfer function.

The transfer of magnetic energy is given by
\begin{equation}
T_M(k) = \int{\hat{\vB}_{\vk} \cdot \nabla \times \widehat{
    \left(\vv \times \vB\right)}_{\vk}^* \,\textrm{d}\Omega_{\vk}},
\label{eq:TM}
\end{equation}
and we can also define the transfer of energy due to the Lorentz force 
\begin{equation}
T_L(k) = \int{\hat{\vv}_{\vk} \cdot \widehat{\left(\vj \times 
    \vB\right)}_{\vk}^* \,\textrm{d}\Omega_{\vk}}.
\end{equation}
Note that this latter term is part of $T_V(k)$; it gives an estimation of 
the alignment between the velocity field and the Lorentz force at each 
Fourier shell (as shown previously in Fig. \ref{fig:jtrans_lowRe}). 
Also, this term represents energy that is transfered from the kinetic 
reservoir to the magnetic reservoir [in the steady state, the integral 
over all $k$ of $T_L(k)$ is equal to the magnetic energy dissipation 
rate, as follows from eq. (\ref{eq:ind})].

\begin{figure*}
\includegraphics[width=16cm]{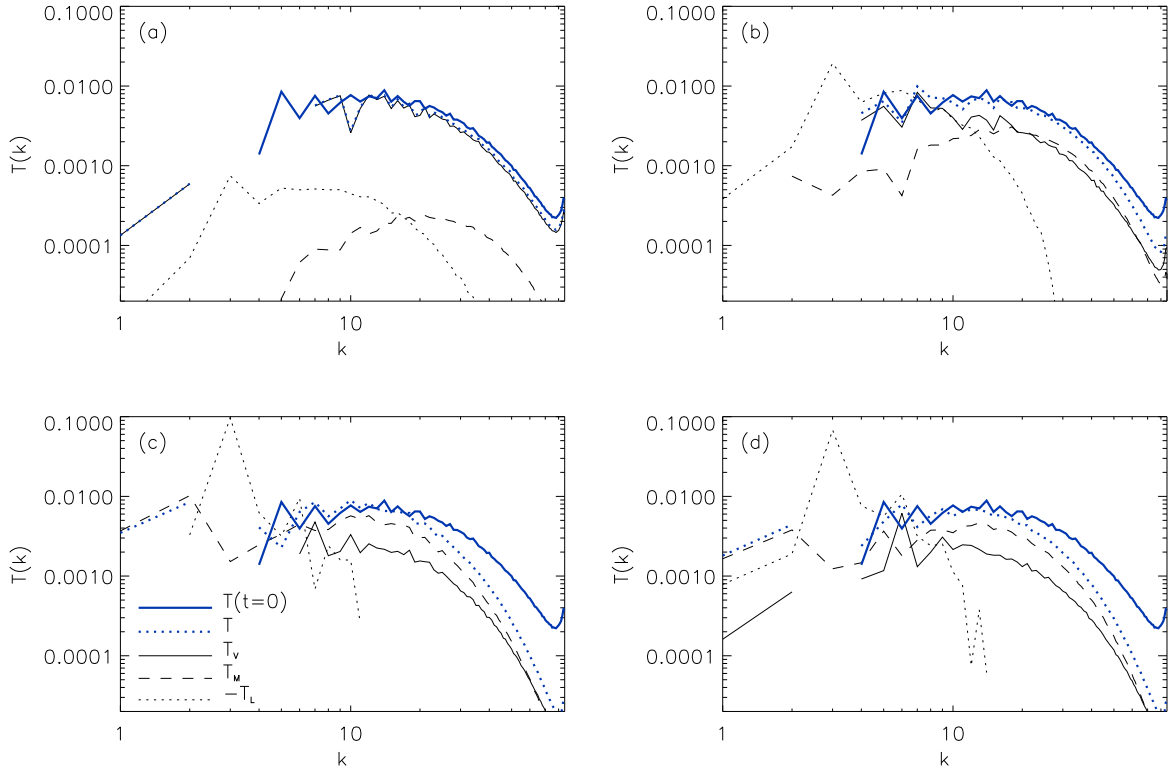}
\caption{(color online) Transfer of energy in Fourier space in run B$'$, at 
         (a) $t=13.5$ in the kinematic regime (see Fig. \ref{fig:Emodes}), 
         (b) $t=73.5$ at the end of the kinematic regime, (c) $t=148.5$ 
         at the time of rapid growth of the $k=2$ shell, and (d) $t=300$. 
         In all the figures, the total transfer at $t=0$ (thick blue solid 
         line) is shown as a reference (which corresponds to the $\vB=0$ 
         case); dotted, dashed, and solid lines represent transfer as 
         indicated in (c). Since $T_L$ is negative at most $k$, $-T_L$ 
         is shown.}
\label{fig:transfer}
\end{figure*}

Fig. \ref{fig:transfer} shows the transfer functions: $T(k;t=0)$ 
(which corresponds to the total energy transfer in the hydrodynamic 
simulation, since the magnetic seed has just been introduced); $T(k)$ 
(which is the total energy transfer); $T_V(k)$; $T_M(k)$; and $-T_L(k)$, 
as functions of $k$ for four different times for run B$'$. A gap in one 
of the spectra indicates (since the plotting is logarithmic) that it has 
changed sign. It is apparent that the dominant transfer is always in 
the vicinity of the forcing band, although it is quite spread over all 
wavenumbers in the inertial range at all times. It is also apparent that 
at the later times, most of the transfer is magnetic transfer, in which, 
of course, the velocity field must participate (see eq. \ref{eq:TM}).

During the kinematic regime (Fig. \ref{fig:transfer}.a), the kinetic 
energy transfer $T_V(k)$ is almost equal to the total transfer. Note 
that the Lorentz force opposes the velocity field at almost all scales, 
and $-T_L(k)$ is approximately constant between $k \approx 3$ and 
$k \approx 12$; all these modes in the magnetic energy grow with the 
same growth rate (see Fig. \ref{fig:Emodes}). The statistically 
anti-parallel alignment between the Lorentz force and the velocity 
field follows from Lenz's law: the electromagnetic force associated 
with the current induced by the motion of the fluid has to oppose the 
change in the field in order to ensure the conservation of energy. Of 
course, the amplified magnetic field is getting its energy from the 
velocity field. Note that then $-T_L(k)$ can be used as a signature of 
the scale at which the magnetic field extracts energy from the velocity 
field (compare this result with the low $R_V$ case, where $-T_L(k)$ 
peaks at $k=3$ both in the kinematic regime and in the nonlinear stage). 
As its counterpart, the transfer of magnetic energy $T_M(k)$ represents 
both the scales where magnetic field is being created by stretching, and 
the nonlinear transfer of energy to smaller scales. $T_L(k)$ is peaked 
at wavenumbers larger than $T_M(k)$; the magnetic field extracts energy 
from the flow at all scales between $k \approx 3$ and $k \approx 12$, 
and this energy turns into magnetic energy at smaller scales through a 
cascade process.

As time evolves and the magnetic small scales saturate, a peak 
in $-T_L(k)$ grows at $k=3$ (Fig. \ref{fig:transfer}.b). At the same 
time, the transfer of kinetic energy $T_V(k)$ at small scales is 
quenched (compared with Fig. \ref{fig:transfer}.a, it has diminished 
in amplitude by almost one order of magnitude). This time corresponds 
to the time where a large scale ($k=2$) magnetic field starts to grow 
(see Figs. \ref{fig:Emodes} and \ref{fig:Bspectra}). In the saturated 
regime (Figs. \ref{fig:transfer}.c,d) $T_L(k)$ is negative at large scales 
and peaks strongly at $k=3$ at late times. As previously mentioned, this 
term represents also energy transfered from the kinetic reservoir to 
the magnetic reservoir. As a result, a substantial fraction of the 
injected energy is seen to be transfered to the magnetic field in the 
injection band ($k=3$), and then most of that energy is carried to small 
scales by the magnetic field ($T_M(k)>T_V(k)$ up to the magnetic dissipation 
scale in the steady state, Figs. \ref{fig:transfer}.c,d). A counterpart 
of this dynamic was observed in \citet{Haugen04}, where it was noted 
by examination of global quantities that most of the energy injected 
in the saturated regime of the dynamo is dissipated by the magnetic 
field. Similarly in run B we find that 
$\left< \nu \omega^2 \right>/\left< \eta j^2 \right> \approx 0.4 $ 
at $t=300$. This explains the drop in the kinetic energy spectrum at 
late times (Fig. \ref{fig:Bspectra}). Note also that the transfer functions 
$T_V(k)$ and $T_M(k)$ drop together at small scales.

In summary, during the kinematic regime the magnetic field is 
amplified in a broad region of $k$-space, while in the nonlinear phase 
most of the amplification takes place at large scales. This contrasts 
with the low $R_V$, $P_M \approx 1$ case, where the magnetic energy 
grows at large scales ($k=2$) from the beginning of the kinematic 
dynamo phase, the small scales being undeveloped.

\begin{figure*}
\includegraphics[width=16cm]{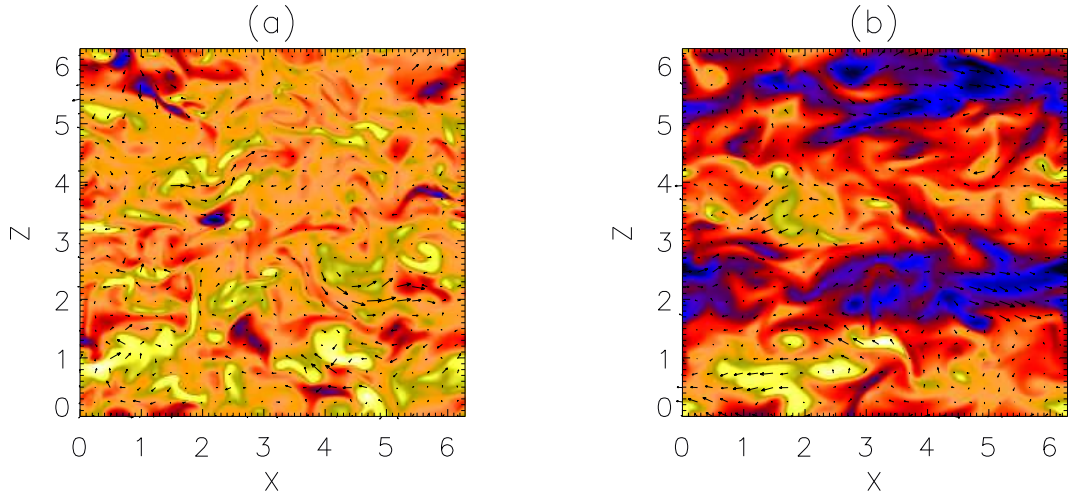}
\caption{(color online) Cross-sectional plots of the magnetic field 
         in the plane $y=\pi/4$ at (a) $t=70$, and (b) $t=300$ 
         ($B_x$, $B_z$ indicated by the arrows, and $B_y$ by the 
         color) for run B$'$.}
\label{fig:highRcut}
\end{figure*}

Finally, we may ask if anything remains visible of any pattern enforced by 
the forcing function $F \vv_{TG}$ in the higher Reynolds number runs. 
Fig. \ref{fig:highRcut} suggests that the answer is yes. It is a plot 
for run B$'$ of a cross section ($y = \pi/4$) in which the magnetic 
field strength is exhibited: arrows denote components in the plane and 
colors denote components normal to the plane. The left panel is at 
$t=70$, and the right panel is at $t=300$. The two horizontal bands 
are associated with the stagnation planes of the Taylor-Green forcing.

While at $t=70$ the magnetic field is mostly at small scale, at 
$t=300$ a pattern reminiscent of the low $R_V$ case 
(Fig. \ref{fig:bcut}.d) can be clearly seen, albeit more turbulent. 
This is the result of the suppression of small scales by the magnetic 
field. While in the kinematic regime the magnetic field grows under 
a broad kinetic energy spectrum, when the large scale magnetic field 
starts to grow the small scale velocity field fluctuations are quenched 
(Figs. \ref{fig:Bspectra} and \ref{fig:transfer}), and the large 
scale pattern reappears. This ``clean up'' effect was also observed 
in \citet{Brandenburg01b}, but there both the large scale pattern 
in the flow and the turbulent fluctuations at small scale were 
imposed, while here the turbulent fluctuations are the result of the 
large scale external force and high values of $R_V$.

\section{\label{sec:discussion}DISCUSSION AND SUMMARY}

The oscillatory behavior exemplified in Fig. \ref{fig:EAB} is not 
without precedent. It is common for a pulsation to occur in a 
faucet when the pressure drop is such as to cause the flow of the 
water to be close to a speed near the threshold of the transition 
to turbulence. The developing turbulence acts as an eddy viscosity 
to reduce the Reynolds number back into the laminar regime. As the 
turbulence then subsides, the flow accelerates until the flow speed 
is again in the unstable regime and the cycle repeats.

A few years ago \citep{Shan93a,Shan93b}, a similarly quasi-periodic 
behavior was observed in an MHD problem which might be considered 
an opposite limit of the dynamo problem. A quiescent, periodic circular 
cylinder of magnetofluid was supported by an external axial magnetic 
field and carried an axial current driven by an applied axial voltage. 
By increasing the axial current, it was possible to cross a stability 
boundary for the onset of mechanical motion. The unstable modes were 
helical, as regards the behavior of $\vv$ and $\vB$. The resulting 
$-\vv \times \vB$ axial electromotive force opposed the sense of the 
applied electric field and constituted an effective increase in the 
resistance of the column. When the disturbances grew large enough, 
the total axial current was reduced back below the stability threshold, 
causing the magnetofluid to re-laminarize itself. A cyclic 
oscillation in magnetic and kinetic energy resulted, with the larger 
energy being magnetic, which in many ways resembles qualitatively the 
oscillations exhibited in Fig. \ref{fig:EAB}, except that the magnetic 
energy remained larger: a sort of ``inverse dynamo'' problem.

In the high Reynolds case, part of this dynamic persists. In the 
steady state of the dynamo, the large scale magnetic field forces 
small scale excitations to be equipartitioned between magnetic 
field and velocity field, and both fields are damped at almost the 
same scale. As a result, velocity fluctuations are strongly 
suppressed and at late times similar structures can be recognized 
in the magnetic field in both the low and high Reynolds simulations.

In all cases, a statistically anti-parallel alignment between the 
Lorentz force and the velocity field was observed by examination of 
nonlinear transfer in Fourier space, stressing the different dynamics 
at different scales. This alignment is more significant at scales 
where the magnetic field is being amplified, and follows from Lenz's 
law applied to a conducting fluid and the conservation of the total 
energy. 

It is clear that there are many distinct dynamo behaviors, depending 
upon the parameters and the nature and scale of the mechanical 
forcing. It should not be inferred that the oscillatory behavior 
shown in Fig. \ref{fig:EAB} is more generic than it is. Different 
kinds of oscillations can appear as the forcing amplitude is varied, 
for example, or as the timing of the seed magnetic field's introduction 
is varied. The oscillations can acquire different qualitative features 
as these features are changed. It may be expected that, once turbulent 
computations in geometries other than rectangular periodic ones are 
undertaken, still further variety may occur.

\acknowledgements

We thank H. Tufo for providing computer time at UC-Boulder, NSF ARI 
grant CDA-9601817. Computer time was also provided by NCAR. The NSF 
grants ATM-0327533 at Dartmouth College and CMG-0327888 at NCAR 
supported this work in part and are gratefully acknowledged.


\end{document}